\documentclass[twocolumn,prd,showpacs,aps]{revtex4}
\usepackage{mathrsfs}
\usepackage{amssymb}
\usepackage{latexsym}
\usepackage{hyperref}
\usepackage{amsmath}
\usepackage{graphicx}

\begin{document}
\title{First law of thermodynamics on holographic screens in entropic force frame}
\author{Yi-Xin Chen}\email{yxchen@zimp.zju.edu.cn}
\author{Jian-Long Li}\email{marryrene@gmail.com}
\affiliation{Zhejiang Institute of Modern Physics, Zhejiang
University, Hangzhou 310027, China}

\begin{abstract}
By following the spirit of Verlinde's entropic force proposal
arXiv:1001.0785, we give the differential and integral form of the
first law of thermodynamics on the holographic screen enclosing a
spherical symmetric black hole. It is consistent with equipartition
principle and the form of Komar mass. The entropy of the holographic
screen determines its area, i. e. $S=\frac{A}{4}$. And then we
express the metric by thermodynamic variables, to give an
illustration of how the space is foliated by the thermodynamical
potentials.

\end{abstract}
 \maketitle

\section{Introduction}
One of the most interesting discovery in general relativity is the
black hole thermodynamics, which gives an implication of the
relation between gravity and thermodynamics \cite{Bardeen}.
Recently, it is wide believed that gravity might be originated from
the thermodynamics of the unknown microstructure of spacetime.
Jacobson first illustrates this idea by obtaining Einstein's
equation from the first law of thermodynamics, $\delta Q= T dS$,
which are defined on a local Rindler horizon. This illustration can
be extended to non-Einstein gravity(for a review, see \cite{10}). An
alternative illustration is the entropic force proposed by Verlinde
\cite{Verlinde:2010hp}. In his work, spacetime is the place where
information is stored, and the fundamental unit of 3-dimensional
space is 2-dimensional holographic screen. By the assumption of
thermodynamical properties on the holographic screens and the
principle of holography, Verlinde gives the correspondence between
gravity in 3-dimensional space and thermodynamics on the
2-dimensional holographic screen. Namely, the origin of the gravity
is the entropic force. Entropic force soon gets various applications
and comments, for an incomplete list, see \cite{Cai:2010hk}.

In Verlinde's proposal, each holographic screen is assigned with a
temperature. The Komar energy of the system enclosed by the screen
is evenly divided over the degrees of freedom of holographic screen
according to the temperature. And gravity is originated by the
entropy variation when a test particle is approaching the screen.
However, Verlinde doesn't give the form of the thermodynamical
entropy. In the work of Padmanabhan \cite{Padmanabhan:2010xh}, Komar
energy can be expressed as equipartition principle, $E=1/2 \int T
dN$. For Einstein theory, we have $dN \sim dA$, and $A$ is the area
of the holographic screen. And for a bifurcation horizon, he gets
$E=2TS$, where $S$ is the entropy on the horizon. Thus we can guess
the entropy of the holographic screen may be $S=A/4$, as it is on
the horizon. In \cite{Lee:2010za, Kiselev:2010mz, Pan:2010eu,
Tian:2010uy}, the authors also get the relation $S=A/4$ from
different aspects in the idea of entropic force, and $S$ is the
entropy on holographic screen. Because $T$, $E$ and $S$ are well
defined on holographic screen, it is necessary for us to study the
first law of the holographic screen thermodynamics.

The black hole horizon is a natural choice of holographic screen, so
it is reasonable to assume that holographic screen thermodynamics
are similar to black hole thermodynamics. Besides, in
\cite{Banerjee:2010yd, Cai:2010prd, Tian:2010uy}, it is found that
when a holographic screen coincides with the black horizon, the
definition of Komar energy is equivalent to the Smarr law, which is
the integrated form of first law of black hole thermodynamics. As a
result, when the holographic screen is far from the horizon, it is
reasonable to assume that the equivalence is still valid. Since
Smarr law is obtained by integrating the differential form of the
first law of black hole thermodynamics, it is required that there
must be a differential first law of holographic screen
thermodynamics.

Moreover, both in Jacobson and Verlinde's work, Unruh temperature
detected by an accelerated observer in each spacetime point plays a
key role. It appears in the first law $\delta Q= T dS$ on the
Rindler horizon and it is equivalent to the holographic screen
temperature. Thus, it is necessary for a corresponding first law of
thermodynamics on the holographic screen. In \cite{Zhao:2010qw},
Zhao discusses the Poincar\'e symmetry of the first law of
thermodynamics. It indicates that thermodynamics as well as gravity
is universal for all the physical system. This is another
implication of the existence of first law of holographic
thermodynamics.

In this article, by using the similar method of Smarr, we give the
first law on the holographic screen enclosing a spherical black
hole, in both the differential and integrated form. The integrated
form is consistent to the energy equipartition principle and the
Komar mass energy. We find that the screen entropy still has the
relation $S \sim A/4$, but we reverse the logic of the
interpretation to $A \sim 4S$. That is, the area of the holographic
screen is determined by its entropy. Comparing to Newton potential,
the holographic screen thermal potential, such as temperature, is
more appropriate for foliating the space. As a result, not only the
Einstein equation is an equation of thermodynamical state
\cite{Jacobson}, but also the space metric can be written as
functions of thermodynamical entities. This a complementary
illustration of the emergency of spacetime by Verlinde.

In section II, we start with a short review of the definition of
temperature and energy from entropic force. The first law of
holographic screen thermodynamics is given in section III. In
section IV, we illustrate our method on an RN black hole. Then we
discuss the physical meaning of $S=\frac{A}{4}$ and how the
thermodynamics generates the space. The conclusion are presented in
section V.

\section{Temperature and energy on holographic screen}
In this section, we review the definition of temperature and energy
on the holographic screen by Verlinde \cite{Verlinde:2010hp}. In the
context we use the units $G=\hbar=c=k_B=1$.

From the perspective of Verlinde, the three-dimensional theory of
gravity as well as the space are originated from the thermodynamics
of the two-dimensional holographic screens covering the
three-dimensional space. The holographic screens are characterized
by temperature and energy, and the `number of bits' from the unknown
microstructure of spacetime.

In the spacetime with a global time-like Killing vector field $\xi$,
we can define the Newton potential
\begin{equation}
\phi =\frac{1}{2} \ln (-\xi^2). \label{Newton_p}
\end{equation}
The holographic screen is defined as an equipotential surface. The
four acceleration of a particle close to the holographic screen is
given by
\begin{equation}
a^\mu =-\nabla^\mu \phi. \label{a}
\end{equation}
Then the holographic screen temperature is defined as the Unruh
temperature for the acceleration,
\begin{equation}
T=-\frac{1}{2\pi}e^\phi n^\mu a_\mu=\frac{1}{2\pi}e^\phi
\sqrt{\nabla^\mu \phi \nabla_\mu \phi}. \label{Temp}
\end{equation}
Here, $n^\mu=\nabla^\mu \phi/\sqrt{\nabla^\nu \phi \nabla_\nu \phi}$
is the normal vector on the screen. The energy of the screen is
defined by the Komar mass energy \cite{Banerjee:2010yd,
Tian:2010uy},
\begin{equation}
E_{Komar}(\mathcal{S},\xi)=\frac{1}{8\pi}\int_{\mathcal{S}}
*d\xi=\frac{1}{2}\int_{\mathcal{S}} T dN. \label{EKomar}
\end{equation}
Here, $N=A$ is the number of bits stored on the holographic screen,
assumed by Verlinde \cite{Verlinde:2010hp}, and $\mathcal{S}$
denotes a holographic screen. The last equation in (\ref{EKomar})
can be interpreted as energy equipartition rule on the screen.

\section{The first law on holographic screen}
In this section, we follow the similar method of Smarr to obtain the
formula of the first law of thermodynamics on holographic screen in
3+1 dimensions, which enclosing a spherical symmetric black hole.
The general metric of this spacetime takes the form,
\begin{equation}
 ds^2 = -f(r) dt^2 + \frac{dr^2}{f(r)} + r^2 d\Omega^2, \label{sph.metric}
\end{equation}
where $d\Omega$ is the line element of an unit 2-sphere. If the
spacetime is asymptotically flat, the function $f(r)$ satisfies,
\begin{equation}
 \lim_{r \rightarrow \infty} f(r) = 1. \label{assume_f(r)}
\end{equation}
When we reach the event horizon $r_0$. we have,
\begin{equation}
f(r)\big|_{r=r_0} = 0. \label{eventH_f(r)}
\end{equation}
As a result, we get the Killing vector, Newton potential, and
Unruh-Verlinde temperature expressed as \cite{Tian:2010uy},
\begin{eqnarray}
 \xi_\mu &=&\left(-f(r),0,0,0\right),\\
 \phi &=&\frac{1}{2}\ln f(r), \label{phi}\\
 T &=& \frac{1}{4\pi}|f'(r)|. \label{T}
\end{eqnarray}
Before talking about the thermodynamics, it should be assumed that
the holographic screens are in a thermal equilibrium. Thus, the
holographic screen is at least an isothermal surface, i. e.
$f'(r)=$const. The Newton potential $\phi$ is also a constant on the
isothermal surface, so we begin with a holographic screen determined
by,
\begin{equation}
f(r) = e^{2\phi}=c. \label{c}
\end{equation}
Here, c is a constant, ranging in $[0,1]$. In
\cite{Verlinde:2010hp}, Verlinde asserts that the amount of coarse
graining for the information on the screens is measured by
$-\frac{\phi}{2}=-\frac{\ln{c}}{4}$. For a black hole, $f(r)$ is
also a function of mass and other thermal entities. Thus, the above
equation can also be expressed as,
\begin{equation}
f(r,M,Q_1,...Q_n) = c. \label{c2}
\end{equation}
Here, \{$Q_n$\} are $n$ conserved charges for the spherical black
hole. Following the similar tricks of Smarr \cite{Smarr}, we solve M
as,
\begin{equation}
M =\mathcal{M}(c,r,Q_1...Q_n). \label{M}
\end{equation}
Differentiating the equation, we get,
\begin{eqnarray}
&dM=\frac{\partial \mathcal{M}(c,r,Q_1...Q_n)}{\partial r} dr+
\sum_i \frac{\partial \mathcal{M}(c,r,Q_1...Q_n)}{\partial Q_i} dQ_i
\\
&=\frac{\partial \mathcal{M}(c,r,Q_1...Q_n)}{\partial r}
\frac{\partial r}{\partial \mathcal{A}} d \mathcal{A}+ \sum_i
\frac{\partial \mathcal{M}(c,r,Q_1...Q_n)}{\partial Q_i}
dQ_i.\label{dM}
\end{eqnarray}
Here, $\mathcal{A}$ is an undetermined area-like function. Eq.
(\ref{dM}) can be perceived as the extended first law of black hole
thermodynamics obtained by Smarr \cite{Smarr}, $dM=T
d\mathcal{A}+\Phi dQ+...$, iff we identify the coefficient of
$d\mathcal{A}$ in Eq. (\ref{dM}) with $T$ determined in Eq.
(\ref{T}). So,
\begin{eqnarray}
\frac{\partial \mathcal{M}}{\partial r} \frac{\partial r}{\partial
\mathcal{A}} = \frac{\partial f}{\partial r}(\frac{\partial
f}{\partial M})^{-1}\frac{\partial r}{\partial \mathcal{A}} =
\frac{1}{4\pi} f'(r)
\end{eqnarray}
and,
\begin{equation}
\mathcal{A}=-4\pi \int(\frac{\partial f}{\partial M})^{-1}
dr.\label{A}
\end{equation}

The area-like function $\mathcal{A}$ is determined by
$f(r,M,Q_1,...Q_n)$. Because the derivative $\frac{\partial
f}{\partial M}$ is determined by the equation of motion,
$\mathcal{A}$ is also determined by the equation of motion, and it
varies in different theories of gravity. In general relativity, for
a spherical black hole, we already know that,
$f(r,M,Q_1,...Q_n)=1-\frac{2M}{r}+g(r,Q_1,...Q_n)$. So, in this
case, $\mathcal{A}=\pi r^2=\frac{A}{4}$, which reproduces the black
hole entropy when $c=0$\footnote{Note that in
\cite{Padmanabhan:2010xh}, the mircoscopic degrees of freedom $dN$
in Eq. (\ref{EKomar}) is determined by $\partial L/\partial
R_{abcd}$, where $L$ is the gravity Lagrangrian and $R_{abcd}$ is
the Riemann tensor. In general relativity, $dN \sim dA$, which is
consistent to our result.}.

Because the metric is spherical symmetric, the other thermodynamical
potentials \{$\Phi_i=\frac{\partial \mathcal{M}}{\partial Q_i}$\}
are also spherical symmetric, and are constant on the holographic
screen. So, it is reasonable for us to treat the holographic screen
as in a thermal equilibrium. And the first law of holographic screen
is obtained by rewriting Eq. (\ref{dM}) as,
\begin{eqnarray}
dM= Td\mathcal{A}+\sum_i \Phi_i dQ_i.\label{dM2}
\end{eqnarray}
It is our key conclusion in this section. This formula looks like
the black hole thermodynamics. However, Eq. (\ref{dM2}) is defined
on any holographic screen with $0 \leqslant c \leqslant 1$, while
the black hole thermodynamics is only the case that $c=0$. The first
law of holographic screen thermodynamics describes the variation
between two adjacent thermal states of the screens, or the variation
after an infinitesimal heat flow absorbed by the screens. So, Eq.
(\ref{dM2}) is also consistent with Jacobson's work \cite{Jacobson}.
It indicates that the gravity is not only a dual of the first law on
a local Rindler horizon, which is a null-like surface, but also a
dual of thermodynamics on a holographic screen which is not
null-like.

\section{RN black hole and integrated first law}

A simple case of the above analysis is the screens enclosing an RN
black hole, with an electric charge $Q$ in metric
(\ref{sph.metric}),
\begin{equation}
f(r) = 1-\frac{2M}{r}+\frac{Q^2}{r^2}. \label{RNmetric}
\end{equation}
Setting $f(r)=c$, we will get two solutions of $r$, and we only take
the solution of the screens which are always lying outside the event
horizon,
\begin{equation}
r_s = \frac{M+\sqrt{M^2-(1-c)Q^2}}{1-c}. \label{RNr}
\end{equation}
Solving $M$ in Eq. (\ref{RNr})
\begin{equation}
M=\frac{r_s}{2}(1-c+\frac{Q^2}{r_s^2}), \label{RN_M}
\end{equation}
And substituting into Eq. (\ref{dM2}), we have,
\begin{equation}
dM= Td\mathcal{A}+\Phi dQ. \label{RN_law}
\end{equation}
Here $T=\frac{f'(r_s)}{4\pi}=\frac{Mr_s-Q^2}{2 \pi
r_s^3}=\frac{(1-c)r_s^2-Q^2}{4 \pi r_s^3}$, $\mathcal{A}=\pi r_s^2$,
and $\Phi=\frac{Q}{r_s}$. When $c=0$, it is the first law of
thermodynamics of RN black hole. To obtain the integrated form of
first law, we integrate Eq. (\ref{RN_law}) by a path in the phase
space of the holographic screen,
\begin{equation}
M=\int_{(r=0,Q'=0)}^{(r=r_s,Q'=0)} T d\mathcal{A}+
\int_{(r=r_s,Q'=0)}^{(r=r_s,Q'=Q)}\Phi dQ'. \label{RN_M2}
\end{equation}
The first integration gets $2T\mathcal{A}+\frac{1}{2}\Phi Q$, and
the second integration gets $\frac{1}{2}\Phi Q$. As a result,
\begin{equation}
M=2T\mathcal{A}+\Phi Q. \label{RN_int}
\end{equation}
This is the integrated form of the first law on the holographic
screen, which is equivalent to the Komar mass energy and the
equipartition principle,
\begin{equation}
M-\Phi Q=2T\mathcal{A}=\frac{1}{2}NT=\frac{1}{8\pi}\int_S
*d\xi=E_{Komar}(\mathcal{S},\xi). \label{RN_equiv}
\end{equation}

So far we have illustrated the differential and integrated form of
the first law of holographic screen thermodynamics, with the screen
enclosing an RN black hole. We can see that $\mathcal{A}$ in the
first law plays the same role as the entropy on the holographic
screen, i. e. $S=\mathcal{A}=\frac{A}{4}$. It is consistent with the
black hole entropy and the holographic screen entropy obtained by
different method in \cite{Lee:2010za, Kiselev:2010mz, Pan:2010eu,
Tian:2010uy}. This interpretation of $\mathcal{A}$ might face two
problems. Why it breaks the Bekenstein entropy bound $S\leqslant 2 E
r$ \cite{Verlinde:2010hp}? When a heat flow $\Delta Q= T \Delta S$
is absorbed by the holographic screen, will the total entropy
$S+\Delta S$ exceed a quarter of the area $\pi r^2$?

The answer for the first problem is that the entropy
$S=\mathcal{A}=\frac{A}{4}$ here is defined on the holographic
screen, just like the equipartition energy $E=M-\Phi Q$, and the
temperature T. The entropy $S$ in Bekenstein bound is defined in the
bulk, which is not applied on the holographic screen. The thermal
entities defined on the holographic screen should be the dual of the
gravitational quantities in the bulk. For example, the energy on the
screen is the dual of the gravitational energy in the bulk, the
temperature is the dual of the gravitational acceleration, and what
is the dual of the entropy on the screen? It is the area of the
2-dimensional boundary of the gravitational system.

Also, we have answered the second problem at the same time. Once a
heat flow is absorbed by the screen, the information stored on the
screen (or in the bulk respectively) is increased, and the
2-dimensional space on the screen is enlarged by the same rate. This
is how the space is emergent from the thermodynamics of the unknown
microstructure. When there is no information, there is no space, and
there is no matter either. When the information comes up, the
2-dimensional space is spanned by the amount of the thermodynamical
entropy of the information. Then the amount of matter is increased
by the heat flow $\Delta Q= T \Delta S$. The temperatures on
different holographic screens can be different, so $T_1 \Delta S_1=
T_2 \Delta S_2$. Numerically, we have $T_1 r_1 \Delta r_1= T_2 r_2
\Delta r_2$, where we have defined $r =\sqrt{S/\pi}$. Thus, in the
view of thermodynamics on the screen, there is a third dimension
emergent from different screens with the distance $(\Delta
r_1-\Delta r_2)$. While in the view of gravity in the bulk, the
holographic screens expand to new positions with new areas as the
black hole horizon does in the same process.

Numerically, the above integrated form of first law Eq.
(\ref{RN_equiv}) is the same to Eq. (\ref{RN_M}), because we have
just differentiated it and then integrated it. Notice that Eq.
(\ref{RN_M}) is obtained by $f(r)=c$, which is a component of
spacetime metric in general relativity without considering
thermodynamics. It indicates that there is a deep relationship
between gravity and holographic screen thermodynamics. i. e.
thermodynamics is hidden in gravity, or gravity is a reflection of
the holographic screen thermodynamics. In \cite{Jacobson}, Jacobson
argued that Einstein equation ``is born in the thermodynamic limit
as a relation between thermodynamic variables". So, it is
straightforward to see that, the spacetime metric which is solution
to Einstein equation, should have a counterpart in thermodynamics.

Let us realized the argument specifically. Since $S=\frac{A}{4}$,
the spherical 2-dimensional part of the metric can write as
\begin{equation}
r^2 d\Omega^2=\frac{S}{\pi} d\Omega^2. \label{metric_sph}
\end{equation}
Thus, to integrate the spherical part of the metric is to run over
all degrees of freedom on the holographic screen. However, there is
a freedom of rewriting the radial part of the metric. From $f(r)=c$
and $T=\frac{(1-c)r_s^2-Q^2}{4 \pi r_s^3}$, we have,

\begin{equation}
\frac{1}{f(r)}dr^2=\frac{1}{1-\frac{4TS+\Phi Q}{\sqrt{S/\pi}}}
(d\sqrt{S/\pi})^2=\frac{1}{1-\frac{2M-\Phi Q}{\sqrt{S/\pi}}}
(d\sqrt{S/\pi})^2. \label{metric_radi}
\end{equation}
Here $dS$ means the variation of the entropy between two screens in
adjacent states. It shows the way that how the space is foliated by
the thermodynamical potential. If we set $\Phi =0$, it is the
radical metric outside a Schwarzschild black hole. If we set $\Phi
=0$ and $T =0$, it is the radical metric of a flat space. If we
define a non zero potential $\Phi^\prime$ and a charge $\Lambda$, we
can also get the metric of an AdS RN black hole. So viewed in
thermodynamics, the space is foliated by the respective
thermodynamical potentials other than Newton potential $\phi$.

\section{Conclusion and discussions}

The differential and integrated form first law of holographic
thermodynamics are obtained from the gravity of spherical black
hole. And the entropy of the holographic screen is $S=\frac{A}{4}$,
with the meaning that the area is determined by the amount of
entropy. By expressing the spherical solution of Einstein equation
by thermodynamic variables, we give an illustration of how the
3-dimensional space is emergent from the thermodynamics. It
indicates that the solution (metric) as well as Einstein equation
are born in thermodynamics.

The radical metric of Eq. (\ref{metric_radi}) suffers a drawback
from the arbitrariness of rewriting $f(r)$ as thermodynamical
entities. e. g. $Q^2/r^2$ can also write as $\Phi ^2$. However, we
should notice that the equivalency between the first law of
thermodynamics and the equation of motion of gravity. So, when we
put $f(r)$ in the equation of motion, it should be equivalent to the
first law. Thus we can not arbitrarily express $f(r)$ by
thermodynamical entities, and we hope we will find a way to fix it
carefully in future.

The work is discussed in the spherical black hole case, in which the
thermodynamical potentials are found to be constant on the screen.
If we want to extend this work on axisymmetric case, such as
Kerr-Newman black holes, we will find that the temperature, the
electric potential, the angular velocity and the Newton potential
are constant only on the black hole horizon \cite{Tian:2010uy}. It
means the holographic screens outside the black hole horizon may not
be states of thermal equilibrium. The dual of thermodynamics and
gravity is hidden deeper in this case.

\textbf{Note added}:When we are in the final stage of writing the
manuscript, a paper \cite{Piazza:2010hz} appears in the preprint
archive, which discusses some relevant topics with our discussion in
this paper. By using geometrical method, the author extends the
result of Jacobson to a time-like screen of observers of finite
acceleration. The form of his result is significantly different from
ours, and the relation between his result and our result still needs
to be clarified.

\acknowledgments

We thank Q.J.Cao, K.N.Shao Y.Q.Wang for useful discussions. Li wants
to thank Y. Wang for helpful discussions. The work is supported in
part by the NNSF of China Grant No. 10775116, 973 Program Grant No.
2005CB724508, and ``the Fundamental Research Funds for the Central
Universities". Chen would like to thank the organizer and the
participants of the advanced workshop, ``Dark Energy and Fundamental
Theory" supported by the Special Fund for Theoretical Physics from
the National Natural Science Foundation of China with grant no:
10947203, for stimulating discussions and comments.


\begin{thebibliography}{99}
\bibitem{Bardeen}
J. M. Bardeen, B Carter and S. W. Hawking, Commun.\
Math.\ Phys.\ {\bf 31}, 161 (1973).

\bibitem{Jacobson}
T. Jacobson, Phys. Rev. Lett. 75, 1260 (1995).

\bibitem{10}
T. Padmanabhan, arXiv:0911.5004.


\bibitem{Verlinde:2010hp}
  E.~P.~Verlinde,
  arXiv:1001.0785 [hep-th].

\bibitem{Cai:2010hk}
  R.~G.~Cai, L.~M.~Cao and N.~Ohta,
  Phys.\ Rev.\  D {\bf 81}, 061501 (2010)
  [arXiv:1001.3470 [hep-th]].

  F.~W.~Shu and Y.~Gong,
  arXiv:1001.3237 [gr-qc].

  M.~Li and Y.~Wang,
  Phys.\ Lett.\  B {\bf 687}, 243 (2010)
  [arXiv:1001.4466 [hep-th]].

  C.~Gao,
  Phys.\ Rev.\  D {\bf 81}, 087306 (2010)
  [arXiv:1001.4585 [hep-th]].

  Y.~Wang,
  arXiv:1001.4786 [hep-th].

  Y.~S.~Myung and Y.~W.~Kim,
  arXiv:1002.2292 [hep-th].

  S.~Gao,
  arXiv:1002.2668 [gr-qc].

  S.~Hossenfelder,
  arXiv:1003.1015 [gr-qc].

  J.~Munkhammar,
  arXiv:1003.1262 [hep-th].

  A.~Sheykhi,
  Phys.\ Rev.\  D {\bf 81}, 104011 (2010)
  [arXiv:1004.0627 [gr-qc]].

  H.~Wei,
  arXiv:1005.1445 [gr-qc].

\bibitem{Padmanabhan:2010xh}
  T.~Padmanabhan,
  arXiv:1003.5665 [gr-qc].
  T.~Padmanabhan,
  Mod.\ Phys.\ Lett.\  A {\bf 25}, 1129 (2010)
  [arXiv:0912.3165 [gr-qc]].
  T.~Padmanabhan,
  arXiv:0911.1403 [gr-qc].

\bibitem{Lee:2010za}
  J.~P.~Lee,
  arXiv:1005.1347 [hep-th].

\bibitem{Kiselev:2010mz}
  V.~V.~Kiselev and S.~A.~Timofeev,
  arXiv:1004.3418 [hep-th].

\bibitem{Pan:2010eu}
  Q.~Pan and B.~Wang,
  arXiv:1004.2954 [hep-th].

\bibitem{Banerjee:2010yd}
  R.~Banerjee and B.~R.~Majhi,
  Phys.\ Rev.\  D {\bf 81}, 124006 (2010)
  [arXiv:1003.2312 [gr-qc]].

\bibitem{Cai:2010prd}
  R.~G.~Cai, L.~M.~Cao and N.~Ohta,
  Phys.\ Rev.\  D {\bf 81}, 084012 (2010)
  [arXiv:1002.1136 [hep-th]].

\bibitem{Tian:2010uy}
  Y.~Tian and X.~Wu,
  Phys.\ Rev.\  D {\bf 81}, 104013 (2010)
  [arXiv:1002.1275 [hep-th]].

  Y.~X.~Liu, Y.~Q.~Wang and S.~W.~Wei,
  arXiv:1002.1062 [hep-th].

  R.~A.~Konoplya,
  arXiv:1002.2818 [hep-th].


\bibitem{Zhao:2010qw}
  L.~Zhao,
  arXiv:1002.0488 [hep-th].

\bibitem{Smarr}
  L. Smarr,
  Phys.\ Rev.\ Lett.\ {\bf 30}, 71 (1973) [Erratum-ibid.\ {\bf 30}, 521 (1973)].


\bibitem{Piazza:2010hz}
  F.~Piazza,
  arXiv:1005.5151 [gr-qc].


\end{thebibliography}
\end{document}